\documentclass[onecolumn,A4]{article}
\usepackage[dvips]{graphics}
\usepackage{epsfig}
\usepackage{subfigure}
\usepackage{cite}
\usepackage{anysize}
\begin{document}

\vskip 1cm
\marginsize{3cm}{3cm}{3cm}{1cm}

\begin{center}
{\bf{\huge Performance Studies of Bulk Micromegas of Different Design Parameters}}\\
~\\
Purba Bhattacharya$^{a*}$, Sudeb Bhattacharya$^b$, Nayana Majumdar$^c$, Supratik Mukhopadhyay$^c$, Sandip Sarkar$^c$, Paul Colas$^d$, David Attie$^d$\\
{\em $^a$ School of Physical Sciences, National Institute of Science Education and Research, Jatni, Bhubaneswar - 752005, India}\\
{\em $^b$ Retired Senior Professor, Applied Nuclear Physics Division, Saha Institute of Nuclear Physics, Kolkata - 700064, India}\\
{\em $^c$ Applied Nuclear Physics Division, Saha Institute of Nuclear Physics, Kolkata - 700064, India}\\
{\em $^d$ DSM/IRFU, CEA/Saclay, F-91191 Gif-sur-Yvette CEDEX, France}\\
~\\
~\\
~\\
~\\
~\\
{\bf{\large Abstract}}
\end{center}

The present work involves the comparison of various bulk Micromegas detectors having different design parameters.
Six detectors with amplification gaps of $64,~128,~192,~220 ~\mu\mathrm{m}$ and mesh hole pitch of $63,~78 ~\mu\mathrm{m}$ were tested at room temperature and normal gas pressure.
Two setups were built to evaluate the effect of the variation of the amplification gap and mesh hole pitch on different detector characteristics.
The gain, energy resolution and electron transmission of these Micromegas detectors were measured in Argon-Isobutane (90:10) gas mixture while the measurements of the ion backflow were carried out in P10 gas.
These measured characteristics have been compared in detail to the numerical simulations using the Garfield framework that combines packages such as neBEM, Magboltz and Heed.

\vskip 1.5cm
\begin{flushleft}

\end{flushleft}

\vskip 1.5in
\noindent
{\bf ~$^*$Corresponding Author}: Purba Bhattacharya

E-mail: purba.bhattacharya85@gmail.com

\newpage

\section{Introduction}
\label{sec:intro}
In a Time Projection Chamber (TPC) \cite {TPCHistory}, the ionization, caused by a traversing charged particle drifts towards the endplate where it gets amplified, collected as signal and processed. 
In order to achieve good spatial resolution while maintaining the capability of handling high luminosity, Micro-Pattern Gaseous Detectors (MPGDs) \cite{MPGD} have been proposed to be used instead of Multi Wire Proportional Counter (MWPC) as TPC endplate \cite{TPC1} for the future linear collider e.g. LC \cite{ILC-TPC}.
The Micromegas (MICRO-MEsh GAseous Structure) is one of the MPGDs that is likely to be used as the TPC endplate \cite{TPC2}.

The Micromegas \cite{Micromegas} is a parallel plate device composed of a very thin metallic micromesh, which separates the low field drift region from the high field amplification region.
The important parameters that determine the choice of a particular Micromegas over another are detector gain, gain uniformity, energy and space point resolutions, comfortable operating regime (in terms of voltage settings, signal strengths), stability and ageing characteristics (ion backflow), etc.                
These parameters are known to depend on the geometry of the detector (amplification gap, mesh hole pitch, wire radius), the electrostatic configuration within the detector, the gas composition, etc.
The bulk Micromegas \cite{Bulk} with an amplification gap of $128~\mu\mathrm{m}$ has been considered to be one of the good choices for a readout system in different TPCs due to its performances in terms of gain uniformity, energy and space point resolutions and low ion-feedback\cite{TPC2, BULK128, TPC3, TPC4}.
It is also efficient to pave large readout surfaces with minimum dead zone.
For some experiments involving low pressure operation, Micromegas detectors having larger amplification gaps are more suited \cite{LowPressure1, LowPressure2}.
Bulk Micromegas detectors of a wide range of amplification gaps have been studied for possible application in rare event experiments \cite{T2Knear, Rare1}.

In this work, we will discuss the experimental and numerical studies illustrating the effects of different amplification gaps and mesh hole pitches on the performance of bulk Micromegas.
A systematic comparison of some of the above mentioned detector characteristics has been carried out to weigh out various possibilities and options and guide our choice for specific applications.
A comparison with the numerical simulations obtained using Garfield \cite{Garfield1, Garfield2}, has been performed to verify the mathematical models and confirm our understanding of the physics of the device.

\section{Experimental Setup}
\label{sec:experiment}

Several small size detectors with an active area of $15~\mathrm{{cm}^2}$ have been fabricated at CEA, Saclay, France and tested at SINP, Kolkata, India.
Each of the bulk Micromegas detectors is equipped with a calendered woven micromesh similar to that of the T2K experiment \cite{T2Knear}.
The design parameters of the prototype detectors are compiled in Table \ref{design}.

\begin{table}[h]
\caption{Design parameters of the bulk Micromegas detectors. All detectors have wire diameter of $18~\mu\mathrm{m}$.}\label{design}
\begin{center}
\begin{tabular}{|c|c|c|}
\hline
& Amplification Gap (in $\mu\mathrm{m}$) & Mesh Hole Pitch (in $\mu\mathrm{m}$)\\
\hline
Bulk 180 A & 64 & 63 \\
\hline
Bulk 180 C & 64 & 78 \\
\hline
Bulk 134 & 128 & 63 \\
\hline
Bulk 180 D & 128 & 78 \\
\hline
Bulk 113 & 192 & 63 \\
\hline
Bulk 183 D & 220 & 78 \\
\hline
\end{tabular}
\end{center}
\end{table}

\begin{figure}
\centering
\subfigure[]
{\label{SetupSINP}\includegraphics[height=0.2\textheight]{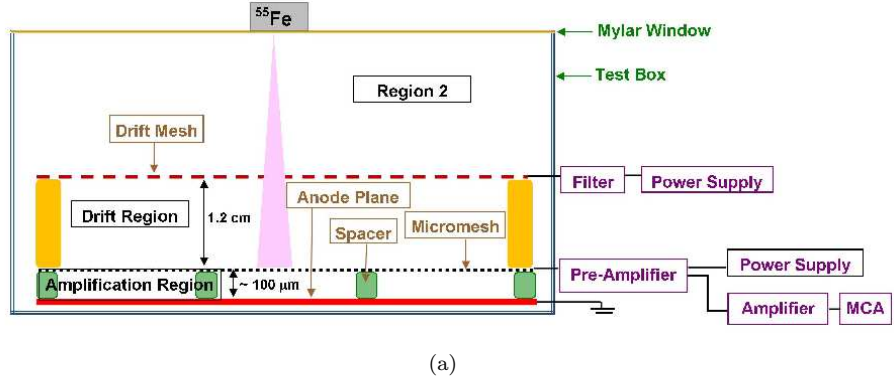}}
\subfigure[]
{\label{SetupSaclay}\includegraphics[height=0.27\textheight]{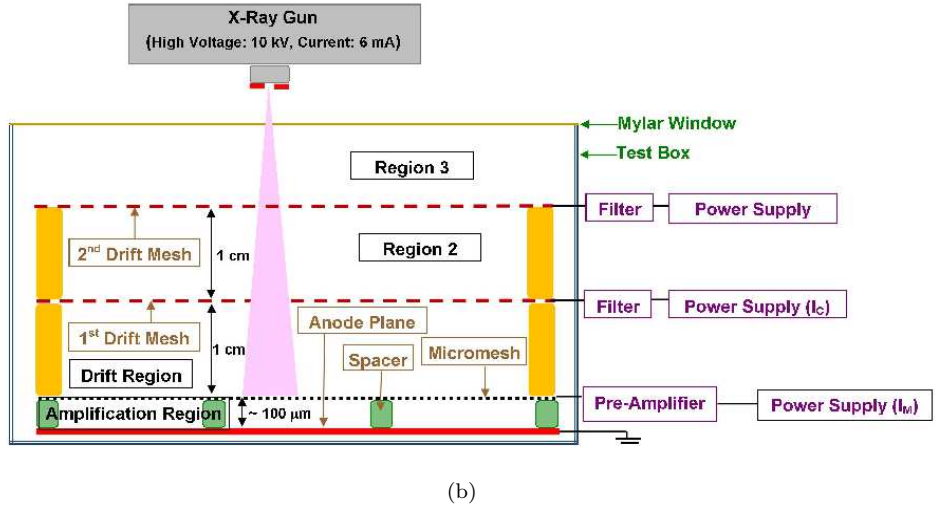}}
\caption{Experimental setup for (a) the measurement of gain, energy resolution and electron transmission; (b) the measurement of ion backflow.}
\label{ExptSetUp}
\end{figure}

The detectors have been characterized by measuring gain, energy resolution, electron transmission using a $^{55}\mathrm{Fe}$ source (present activity 111.56 MBq).
For these studies we had used the setup as shown in Figure \ref{SetupSINP}.
A drift mesh was mounted above the detector.
During the experiment, the drift distance was maintained at 1.2 cm.
The chamber was flushed with Argon-Isobutane gas mixture (90:10) at room temperature (296 K) and atmospheric pressure.
The detector and the drift mesh were biased using CAEN N471A high voltage unit.
The output pulse from the detector was passed through a charge sensitive pre-amplifier (ORTEC model 142IH).
Subsequently, it was fed to a spectroscopic amplifier (ORTEC model 672) with a shaping time constant of $1~\mu\mathrm{sec}$.
Finally, the data were recorded in a multi-channel analyzer (AMTEK MCA 8000A).
The working rate was maintained at $\sim~1200~\mathrm{counts/sec}$.

A simple setup was built at Saclay for the ion backflow measurement as shown in Figure \ref{SetupSaclay}.
While measuring the ion backflow using a single drift mesh, besides the contribution of ions from the avalanche, there was a possibility of having an additional contribution to the drift current from the ions created in Region 2 in Figure \ref{SetupSINP}.
So a second drift mesh was placed at a distance of $1~\mathrm{cm}$ above the first one and kept at the same voltage as that in the first drift mesh.
The ions created between the test box window and the upper drift mesh (Region 3 in Figure \ref{SetupSaclay}) were collected on the outer drift mesh (second Drift Mesh in Figure \ref{SetupSaclay}).
Thus, the current on the inner drift mesh ((first Drift Mesh in Figure \ref{SetupSaclay}) was chosen as an estimate of the ionic current from the avalanche.
The measurements were performed using an X-ray tube (model XRG 3000, operating high voltage: 10 kV, current: 6mA) to produce the primary electrons in the drift region (Figure \ref{SetupSaclay}).
The currents on the drift mesh (first Drift Mesh in Figure \ref{SetupSaclay}) and the micromesh were measured from the high voltage power supply.

\section{Simulation Tools}
\label{sec:simulation}

The experimental data have been compared with the estimates obtained through numerical simulation.
We have used the Garfield (not Garfield$++$) simulation framework.
This framework was augmented in 2009 through the addition of the neBEM \cite{neBEM1, neBEM2, neBEM3, neBEM4, neBEM5} (nearly exact Boundary Element Method, which is known to be very accurate throughout the computational domain) toolkit to carry out 3D electrostatic field simulation.
Earlier, Garfield had to import the field-maps from one of the several commercial FEM (Finite Element Method) packages in order to study the 3D gas detectors.
Besides neBEM, the Garfield framework provides interfaces to HEED \cite{HEED1, HEED2} for the primary ionization calculation and Magboltz \cite{Magboltz1, Magboltz2} for computing the drift, diffusion, Townsend and attachment coefficients.
A short description of the Garfield routines that have been used for numerical simulation is given in the Table \ref{command}.

\begin{table}[h]
\caption{Description of the Garfield routines}\label{command}
\begin{center}
\begin{tabular}{|c|c|c|}
\hline
Name of the routines & Description of the routines & Where the routines are used\\
\hline
$\mathrm{Drift}\_\mathrm{Microscopic}\_\mathrm{Electron}$ & The electron tracking at the & To calculate the electron \\
& molecular level using Monte & transmission ($\eta$) in \\
& Carlo technique. A typical & section in \ref{sec: transparency}, \ref{sec: gain} and \ref{sec: energy resolution}\\
& drift path proceeds through & \\
& millions of collisions. Each & \\
& collision is classified as & \\
& either elastic, inelastic, & \\
& attachment, ionisation  etc & \\
& according to the relative & \\
& cross sections of these & \\
& processes. & \\
\hline
\hline
$\mathrm{Drift}\_\mathrm{Electron}\_\mathrm{3}$ & Using Runge Kutta Fehlberg & To calculate the multiplication \\
& method calculates a electron drift & factor ($\mathrm{g}_{\mathrm{mult}}$) in section \ref{sec: gain} \\
& line and returns the drift time, & \\ 
& multiplication and losses & \\
& through attachment. & \\
\hline
\hline
$\mathrm{Drift}\_\mathrm{MC}\_\mathrm{Ion}$ & Performs Monte Carlo calculation & To calculate the ion drift \\
& of a drift line for a & path which is needed \\
& positively charged ion, taking & in the ion backflow\\
& diffusion into account on a & calculation in section \ref{sec: ion back flow}\\
& step-by-step basis & \\
\hline
\hline
Avalanche & Simulates an electron induced & To calculate the electron\\
& avalanche, taking diffusion, & avalanche which is needed \\
& multiplication, attachment into  & for the gain variation in section \\
& account. The drifting of electrons & \ref{sec: energy resolution} and the ion backflow in \\
& is performed using Monte Carlo & section \ref{sec: ion back flow}.\\
& technique. In comparison to & \\
& $\mathrm{Drift}\_\mathrm{Microscopic}\_\mathrm{Electron}$ & \\
& the step size has to be set & \\
& properly & \\
\hline
\end{tabular}
\end{center}
\end{table}

\section{Results}
\label{sec: results}

\subsection{Electric Field}
\label{sec: field}

Figure \ref{FieldPitch} shows the axial electric field for two different detectors having the same amplification gap, but two different mesh hole pitches.
For the same mesh voltage, the amplification field is lower in case of the larger pitch.
For a particular hole opening and for the same mesh voltage, the variation of the amplification gap changes the axial amplification field as shown in Figure {\ref{FieldGap}.
The effect of these geometrical parameters on the transverse field is shown in Figures  \ref{FieldXPitch} and \ref{FieldXGap}.

\subsection{Electron Transmission of Micromesh}
\label{sec: transparency}

Experimentally we estimate the electron transmission of the micromesh as the ratio of the signal amplitude at a given drift field over the signal amplitude at the drift field where the signal amplitude is maximum.
The measured values and simulated estimates of the electron transmission are compared in Figure \ref{Transparency-Figure}.
These measurements are sensitive to the electron transport at the micron-scale, and thus we use the microscopic electron tracking method, available in Garfield, for the numerical simulation \cite{Garfield1, Transparency1}.
The electrons are injected on square areas with sides equal to that of the size of the pitch.
The first area is placed $100~\mu\mathrm{m}$ above the mesh and subsequent areas are placed from $1~\mathrm{mm}$ to $1.2~\mathrm{cm}$ with a spacing of $1~\mathrm{mm}$ between them. 
Each square is subdivided into $10\times10$ grid pattern. 
Each grid on the square is injected with an electron 100 times.
As a result, $100\times100\times13$ electrons are made to drift towards the amplification region.

\begin{figure}
\centering
\subfigure[]
{\label{FieldPitch}\includegraphics[height=0.2\textheight]{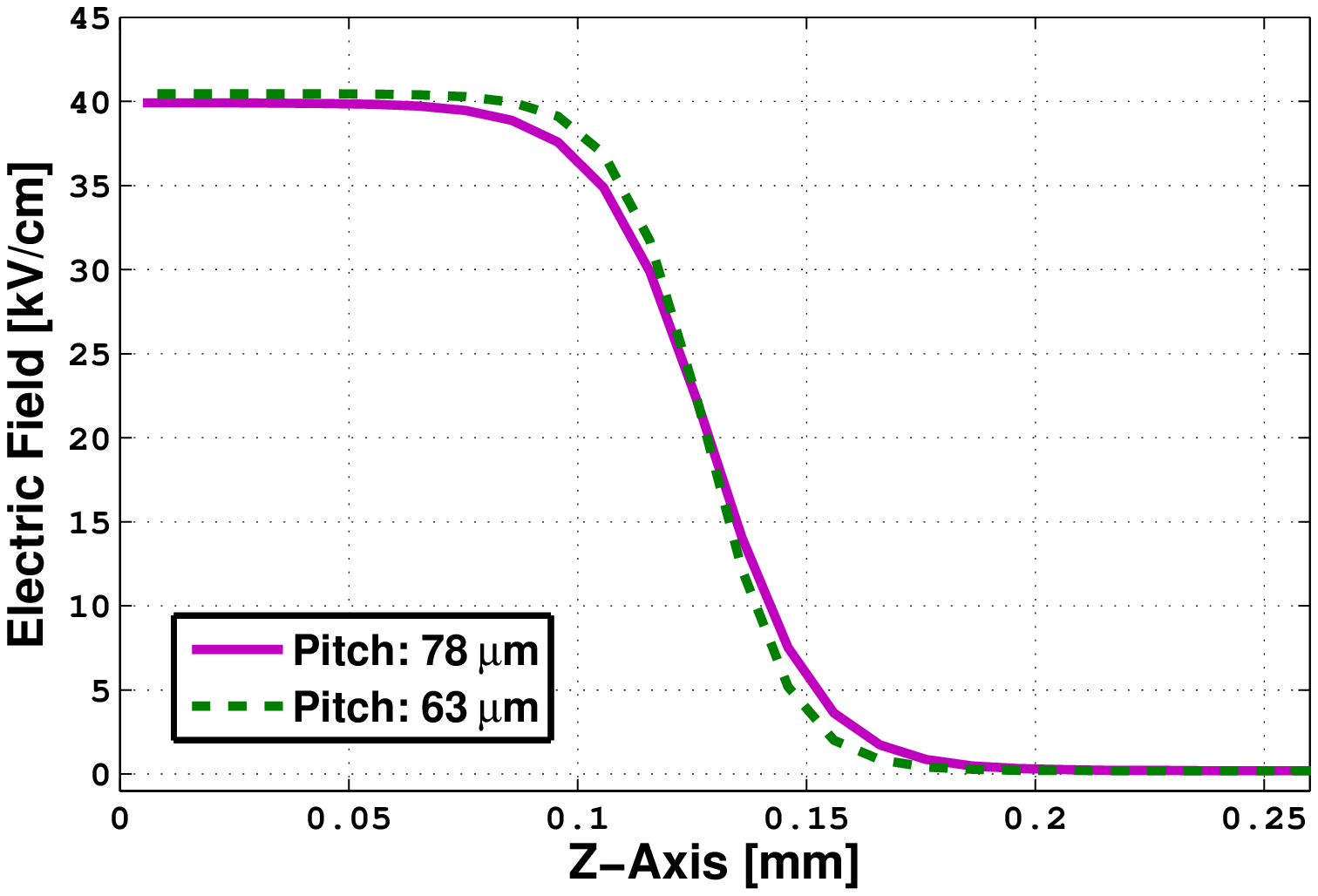}}\llap{\raisebox{2.8cm}{\includegraphics[height=2.3cm]{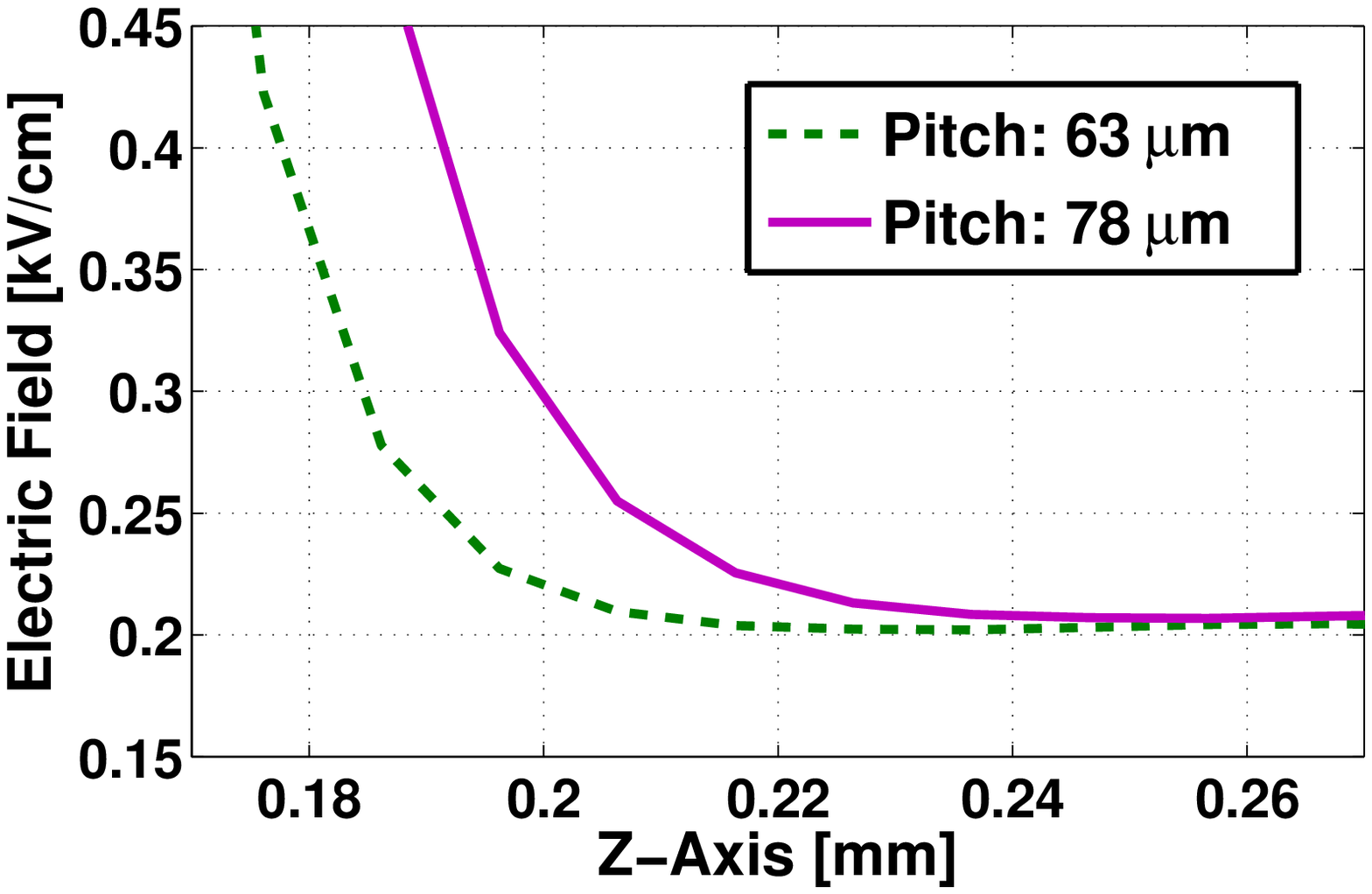}}}
\subfigure[]
{\label{FieldGap}\includegraphics[height=0.2\textheight]{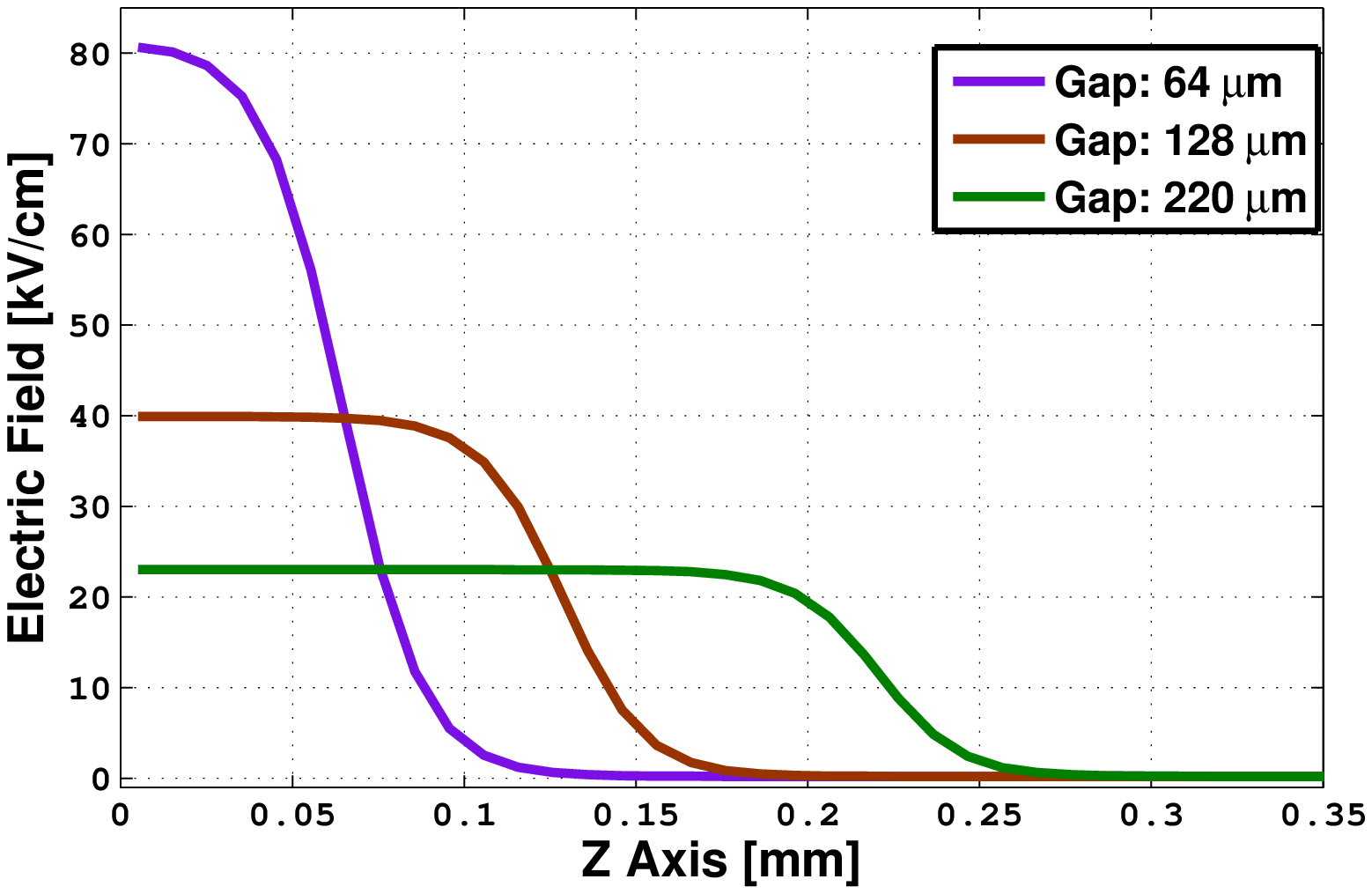}}
\subfigure[]
{\label{FieldXPitch}\includegraphics[height=0.2\textheight]{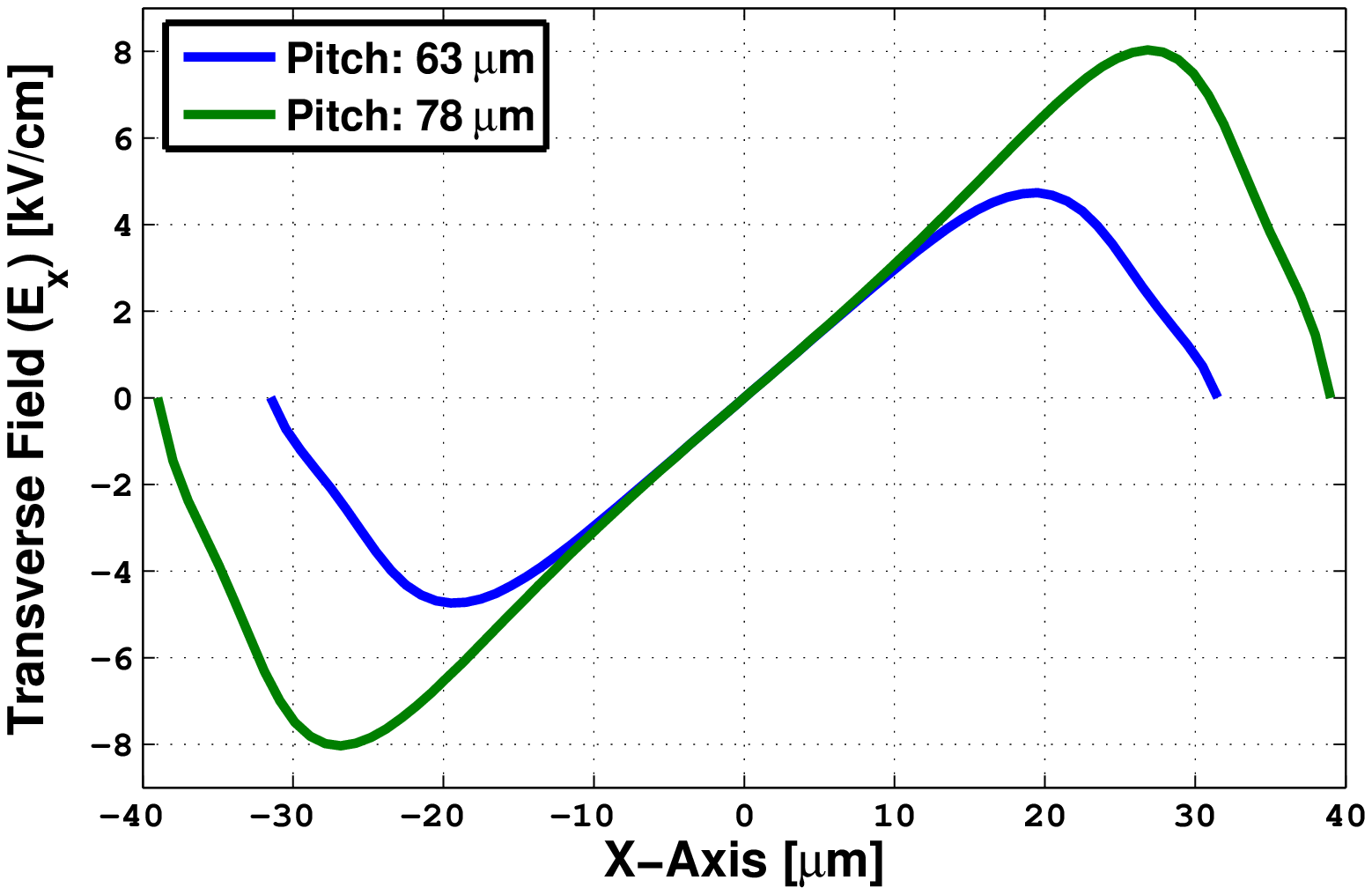}}
\subfigure[]
{\label{FieldXGap}\includegraphics[height=0.2\textheight]{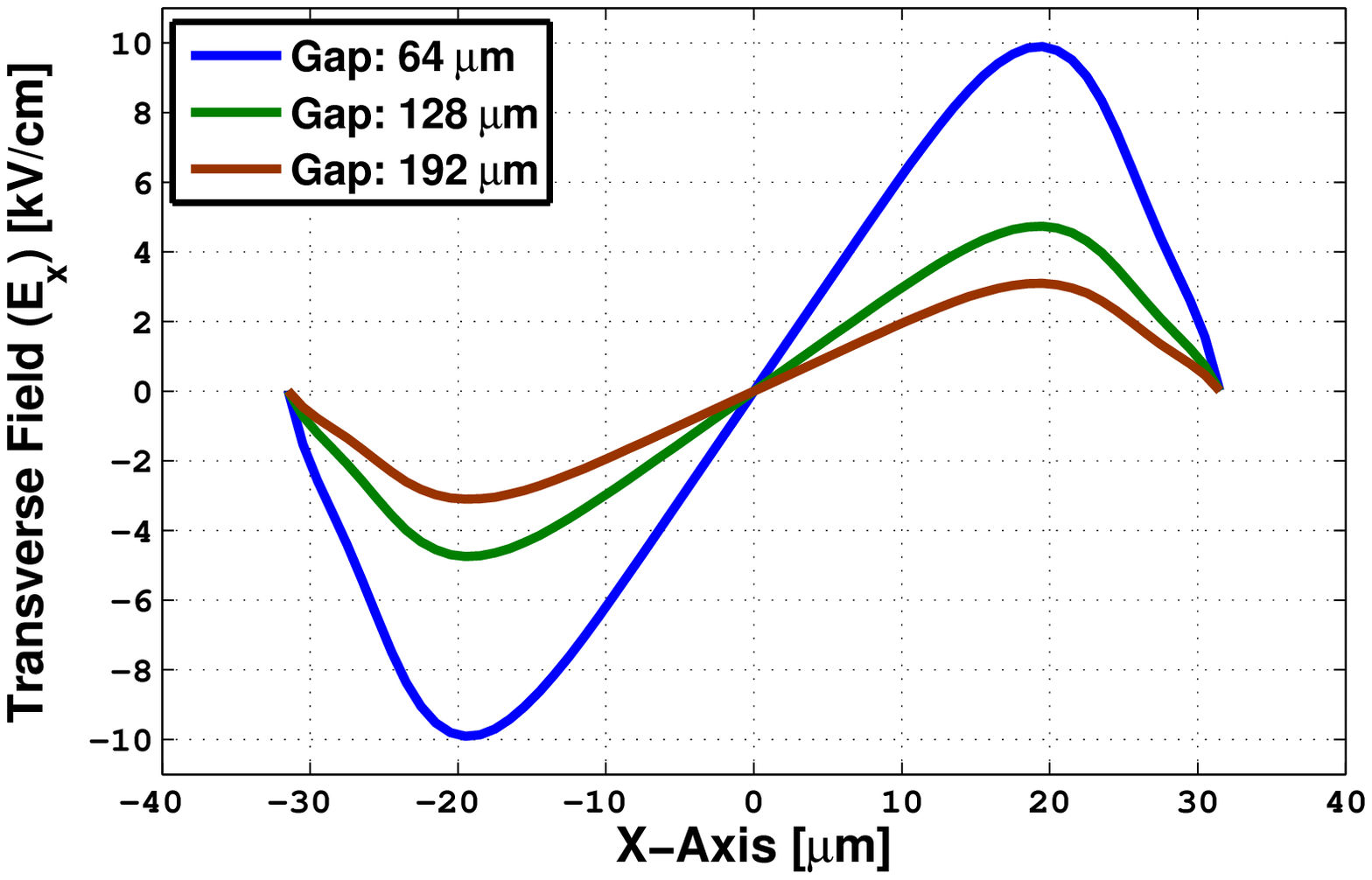}}
\caption{The axial electric field for (a) different mesh hole pitches having the same amplification gap of $128~\mu\mathrm{m}$; (b) different amplification gaps having the same mesh hole pitch of $78~\mu\mathrm{m}$. The transverse electric field close to the hole entrance ($1~\mu\mathrm{m}$ above the micromesh) for (c) different mesh hole pitches having the same amplification gap of $128~\mu\mathrm{m}$; (d) different amplification gaps having the same mesh hole pitch. In each of these cases, mesh voltage = $-500~\mathrm{V}$, drift field = $200~\mathrm{V/cm}$. An expanded view of the drift field is shown in the inset}
\label{Field-EZ}
\end{figure}

The trend of both the experimental and simulation results (at the drift field range of 200 - 2000 V/cm) for the Micromegas with $128~\mu\mathrm{m}$ gap and $63~\mu\mathrm{m}$ pitch (Figure \ref{Trans-Pitch}) are similar to the results of reference \cite{Transparency1}. 
The transmission is maximum between the drift field of 200 V/cm to 750 V/cm.
At higher drift field, the electrons increasingly often hit the mesh and the transmission decreases.                                          
At lower drift field (25 - 100 V/cm), the electron attachment is likely to be responsible for less transmission.
In the simulation, introduction of an impurity of $0.05\%$ Oxygen has been found to reproduce a similar effect.
As an example, for the Micromegas with $128~\mu\mathrm{m}$ gap and $63~\mu\mathrm{m}$ pitch, at the drift field of 400 V/cm, this level of impurity reduces the maximum transmission from $99.7~\%$ to $93.1~\%$ with respect to the initial number of electrons released in the drift volume.
Further detailed work is being carried out in this area.

Figure \ref{Trans-Pitch} shows the variation of the electron transmission with the drift field for two different mesh hole pitches.
For the larger pitch, the maximum number of electrons reach the amplification gap at a higher drift field.
The drift field at which the maximum transmission is achieved also depends on the amplification gap as shown in Figure \ref{Trans-Gap}.
For a particular pitch, the maximum transmission is achieved at a lower drift field when the amplification gap increases.
In each of the cases, the simulation results agree quite well with the experimental data even at a drift field as low as $25~\mathrm{V/cm}$.

\begin{figure}[hbt]
\centering
\subfigure[]
{\label{Trans-Pitch}\includegraphics[height=0.20\textheight]{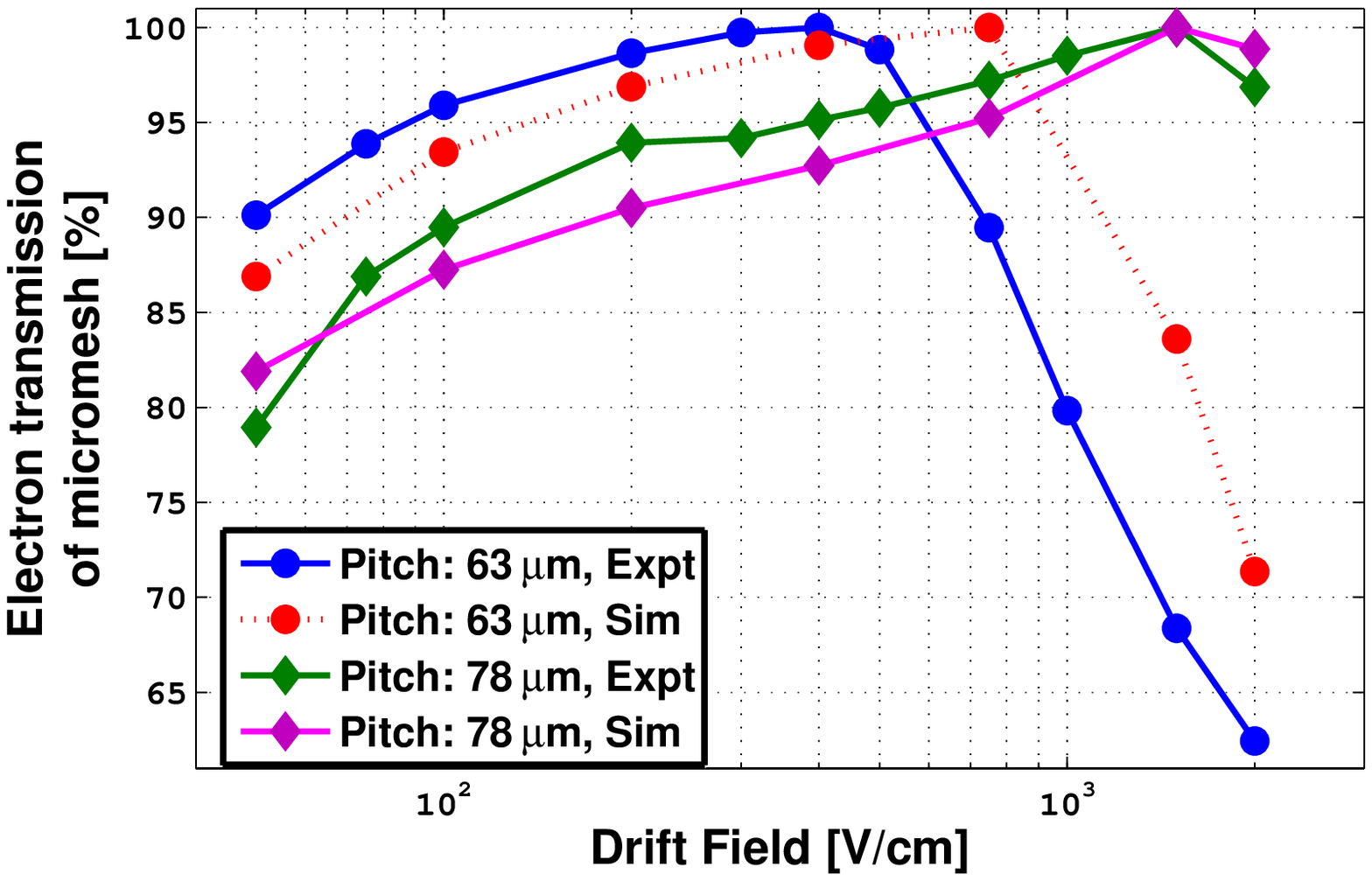}}
\subfigure[]
{\label{Trans-Gap}\includegraphics[height=0.20\textheight]{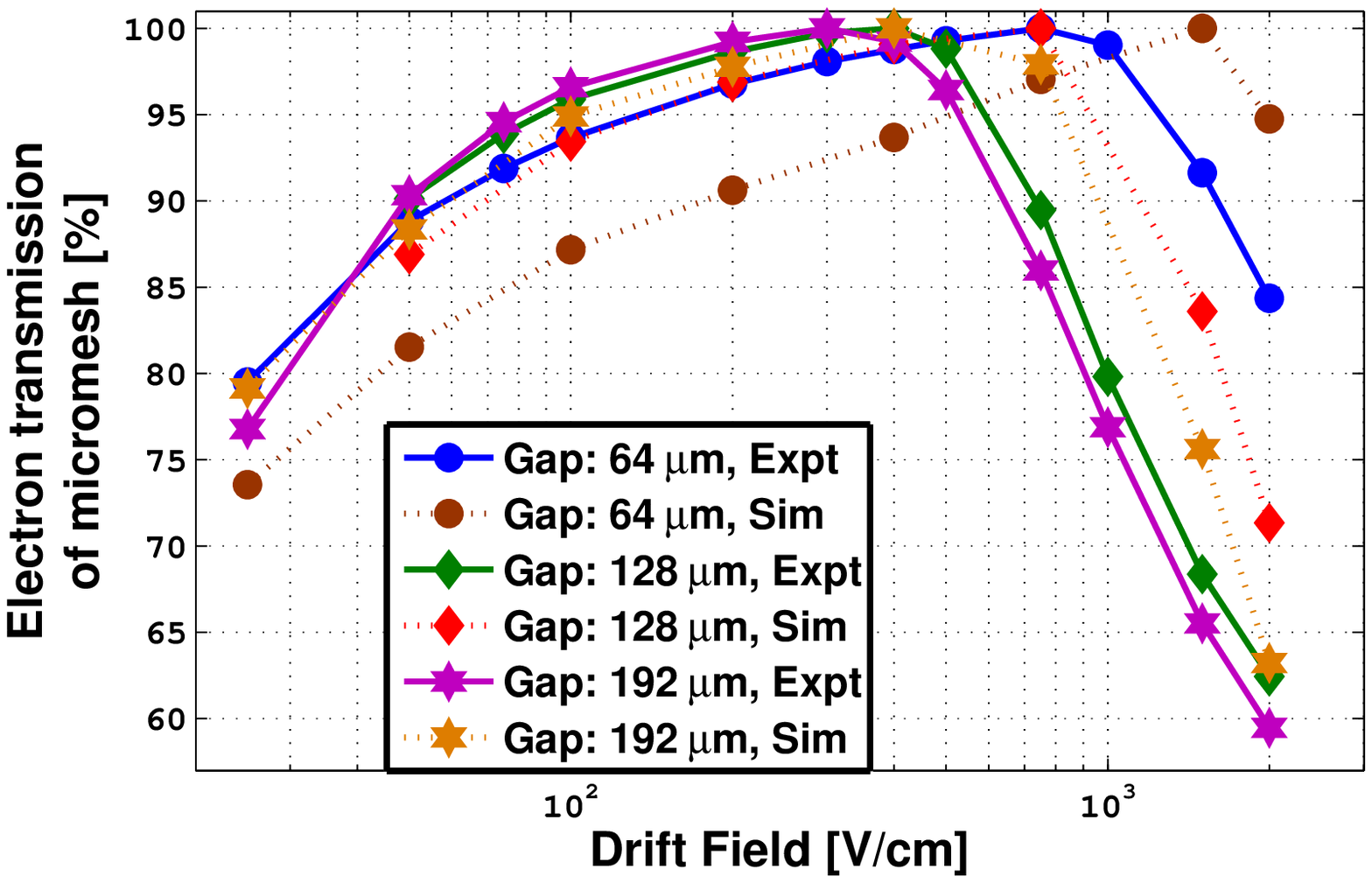}}
\subfigure[]
{\label{Trans-Pitch-1}\includegraphics[height=0.20\textheight]{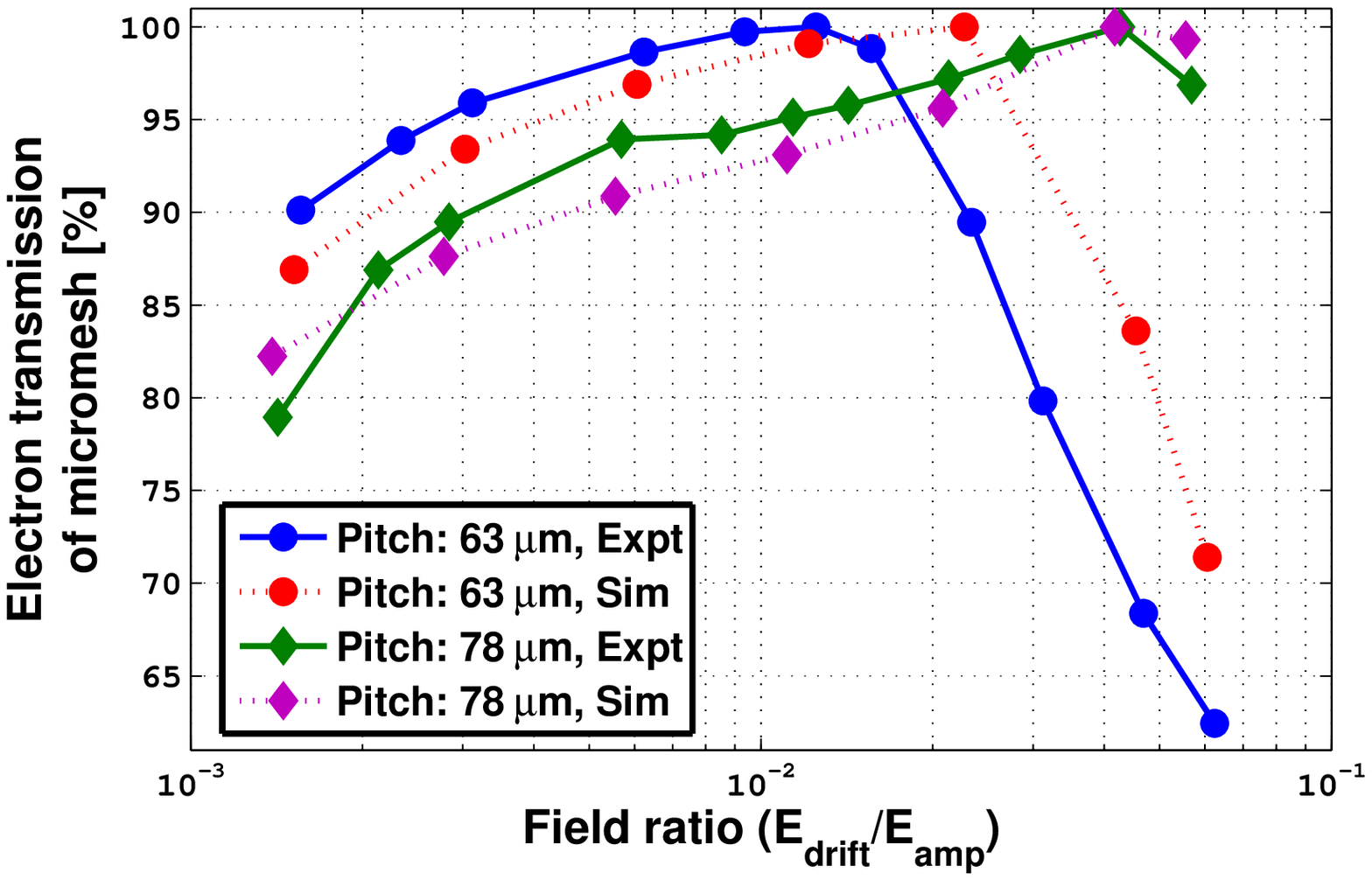}}
\subfigure[]
{\label{Trans-Gap-1}\includegraphics[height=0.20\textheight]{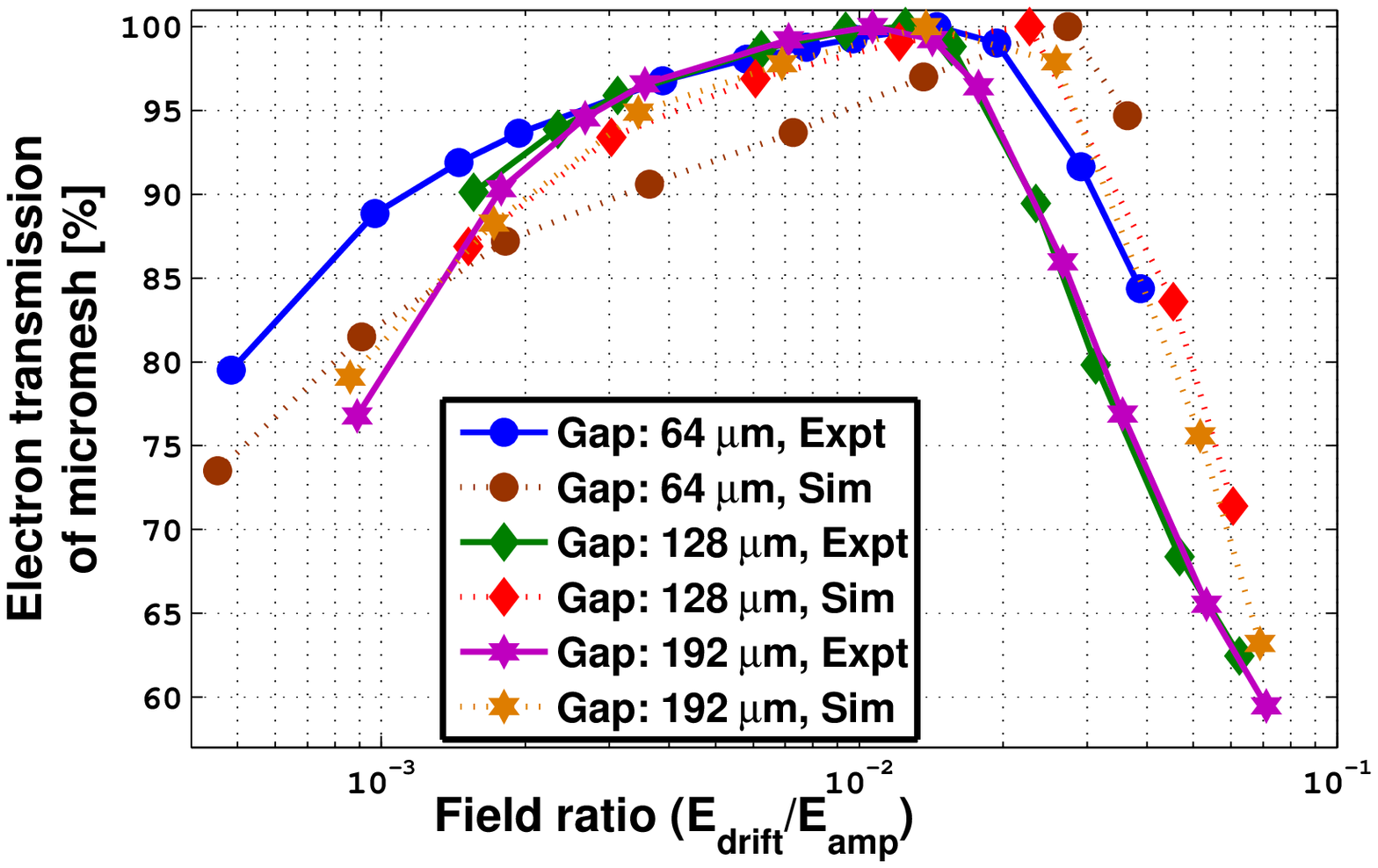}}
\caption{Variation of the electron transmission of the micromesh with the drift field in  Argon-Isobutane mixture (90:10) for (a) two different mesh hole pitches [pitch = $63~\mu\mathrm{m}$, mesh voltage = $-410~\mathrm{V}$, gain $\sim~3200$; pitch = $78~\mu\mathrm{m}$, mesh voltage = $-450~\mathrm{V}$, gain $\sim~4000$] having the same amplification gap of $128~\mu\mathrm{m}$; (b) three different amplification gaps [gap = $64~\mu\mathrm{m}$, mesh voltage = $-330~\mathrm{V}$, gain $\sim~3000$; gap = $128~\mu\mathrm{m}$, mesh voltage = $-410~\mathrm{V}$, gain $\sim~3200$; gap = $192~\mu\mathrm{m}$, mesh voltage = $-540~\mathrm{V}$, gain $\sim~3600$] having the same mesh hole pitch of $63~\mu\mathrm{m}$. (c) and (d) represent the same variation but with the field ratio ($\mathrm{E}_{\mathrm{drift}}/\mathrm{E}_{\mathrm{amp}}$).}
\label{Transparency-Figure}
\end{figure}

\subsection{Gain}
\label{sec: gain}

The measured and simulated gain curves for different bulk Micromegas detectors in an Argon-Isobutane gas mixture (90:10) are presented in Figure \ref{Gain-Figure}. 
Figure \ref{Gain-Pitch1} shows the variation of the gain with the mesh voltage for two different detectors having different pitches.
Our study of the electric field indicates, for the larger pitch, a higher voltage is needed to obtain the same gain.
A comparison between different amplification gaps reveals that the maximum gain achieved experimentally with a larger gap is more than that with a smaller gap (Figure \ref{Gain-Gap1}).
In each of these cases, the numerical results follow the experimental trend.
For the numerical simulation, the effective gain \cite{Gain} is obtained as 
\begin{eqnarray}
\mathrm{g}_\mathrm{eff} = \eta \times \mathrm{g}_\mathrm{mult} 
\end{eqnarray}

\noindent where $\eta$ is the probability for a primary electron to reach the amplification region i.e the electron transmission and $\mathrm{g}_\mathrm{mult}$ is the multiplication factor of the electrons throughout their trajectories. 
Electron transmission and multiplication factor have been computed as a two step approach as outlined in reference \cite{Gain} and reflected in Table \ref{command}.
The Argon-based gas mixtures are Penning mixtures.
After considering the results using different transfer rates, we chose $40 - 50\%$ transfer rate for Argon-Isobutane mixture (90:10) for our calculation. 
This transfer rate is slightly higher than that estimated ($40\%$) in reference \cite{Penning}  and needs further investigation.  

\begin{figure}[hbt]
\centering
\subfigure[]
{\label{Gain-Pitch1}\includegraphics[height=0.2\textheight]{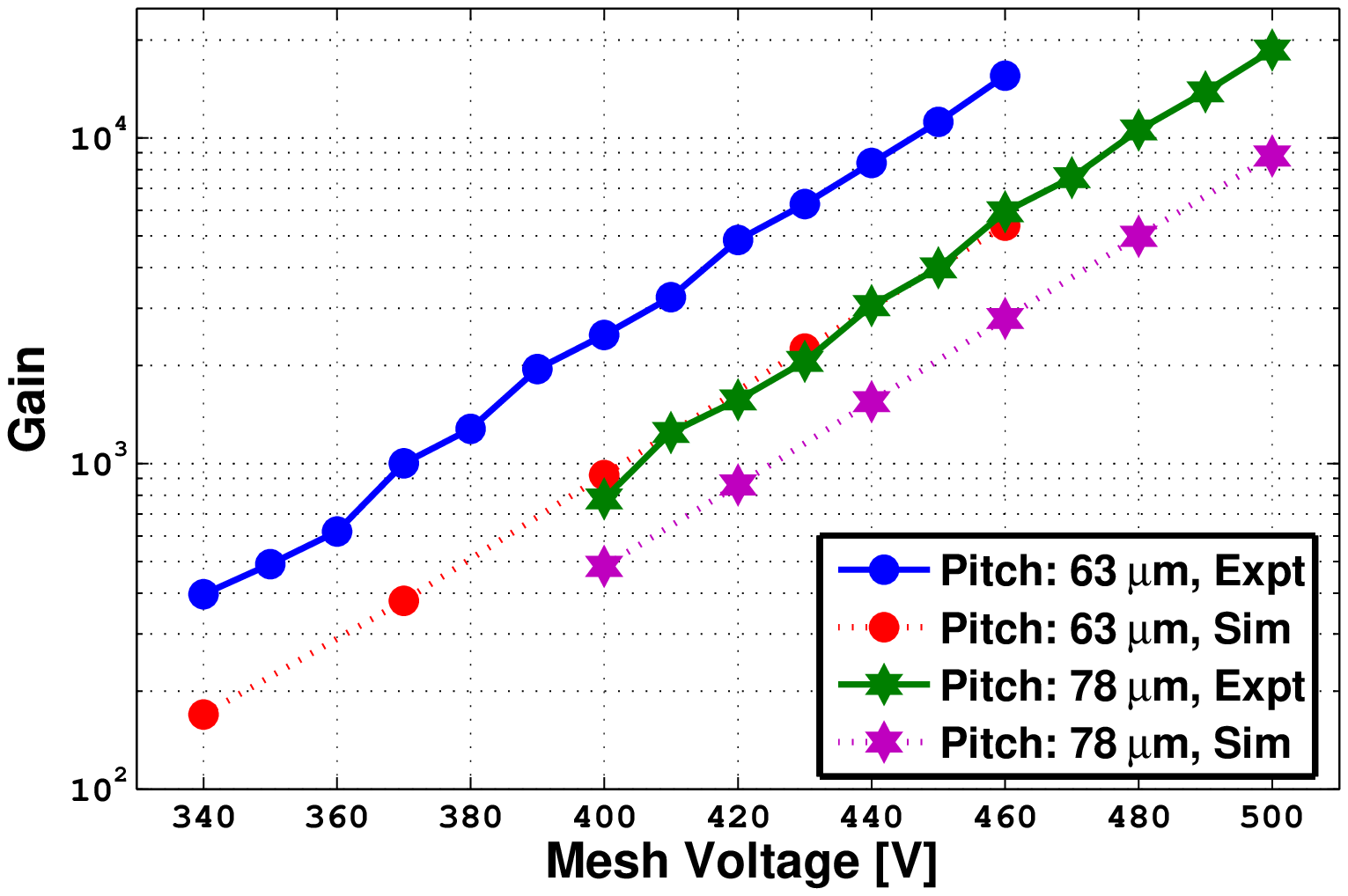}}
\subfigure[]
{\label{Gain-Gap1}\includegraphics[height=0.2\textheight]{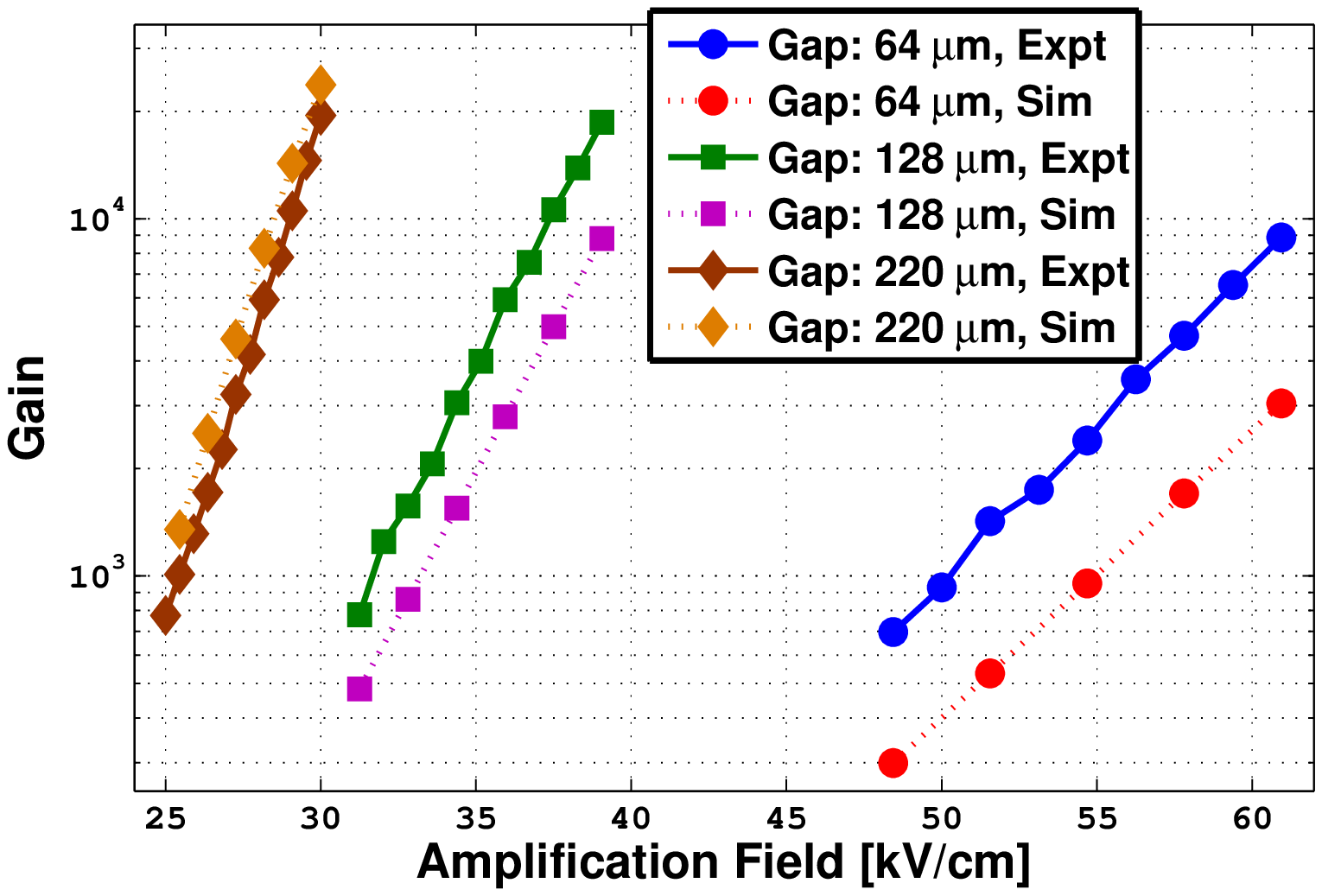}}
\subfigure[]
{\label{Gain-Pitch2}\includegraphics[height=0.2\textheight]{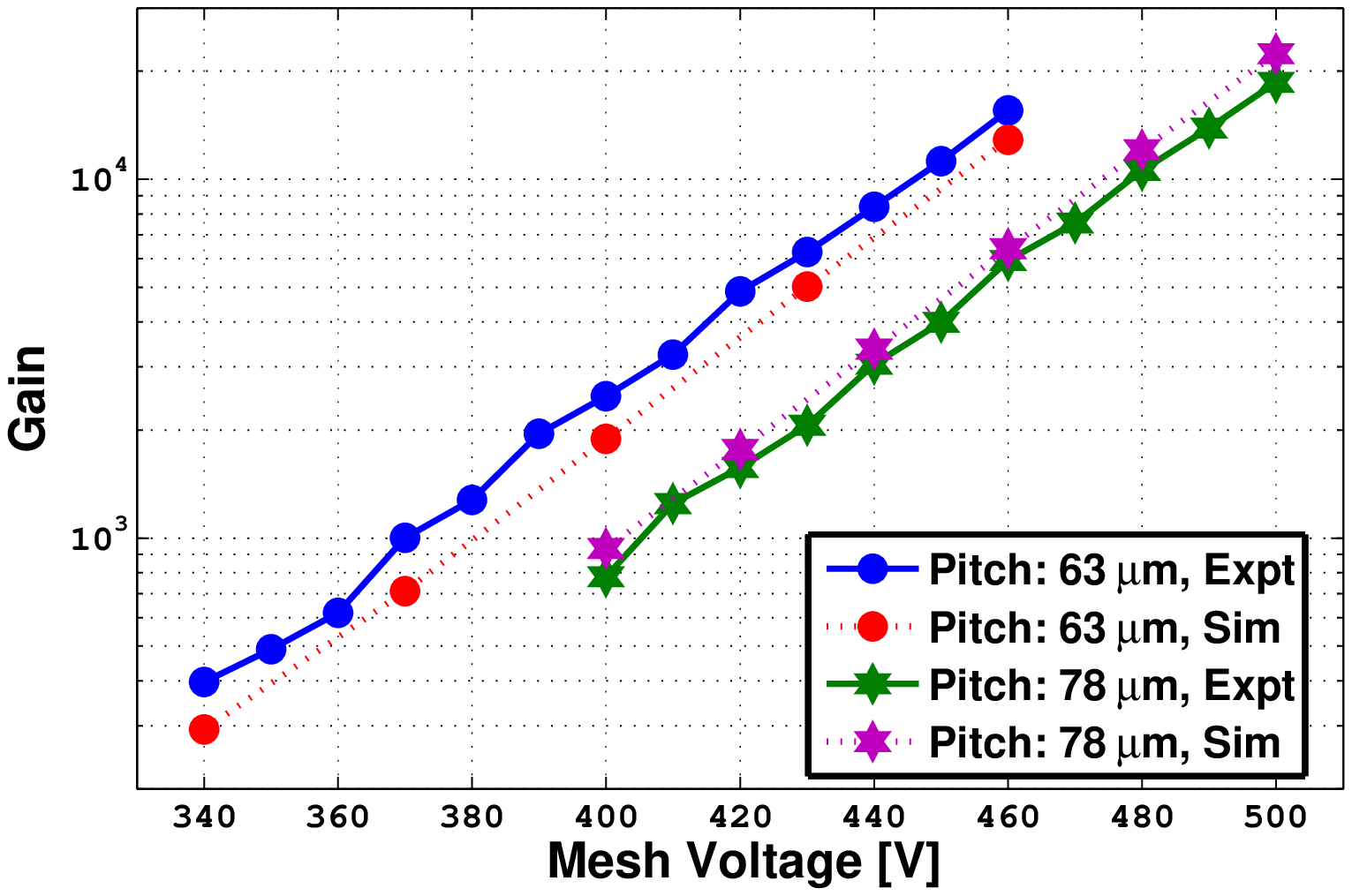}}
\subfigure[]
{\label{Gain-Gap2}\includegraphics[height=0.2\textheight]{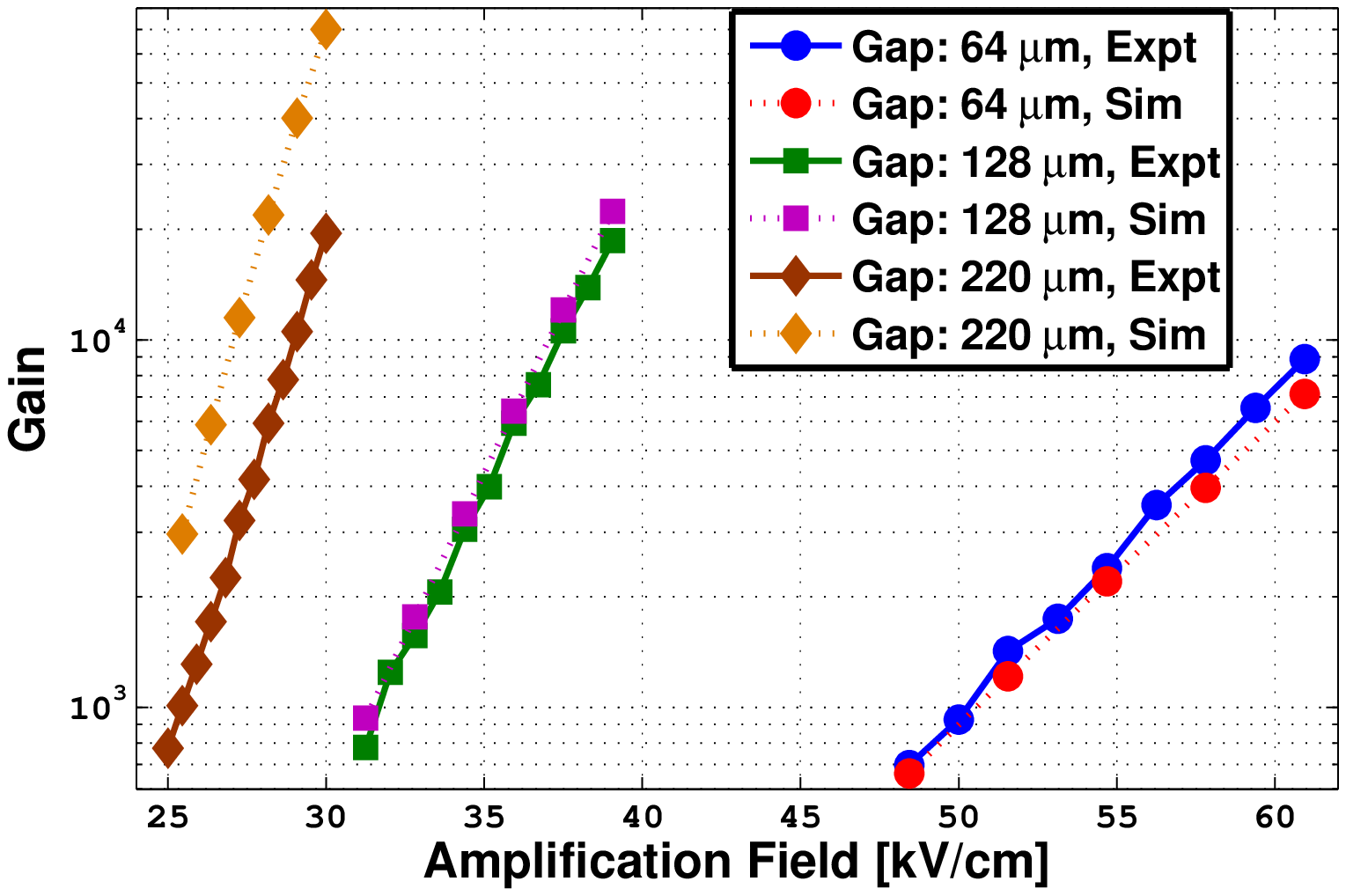}}
\caption{Variation of gain in Argon-Isobutane mixture (90:10) with (a) and (c) mesh voltage for two different mesh hole pitches having the same amplification gap of $128~\mu\mathrm{m}$; (b) and (d) amplification field for three different amplification gaps, having same the mesh hole pitch of $78~\mu\mathrm{m}$. In each of these cases the drift field is $200~\mathrm{V/cm}$. In (a) and (b) the simulation has been performed using $40\%$ Penning. In (c) and (d) the same variation of gain has been shown but using $50\%$ Penning in simulation.}
\label{Gain-Figure}
\end{figure}

\subsection{Energy Resolution}
\label{sec: energy resolution}

The dependency of the energy resolution on the drift field for two different mesh hole pitches and three different amplification gaps is shown in Figure \ref{Energy-Pitch} and Figure \ref{Energy-Gap} respectively. 
At a higher drift field, the detectors with larger pitch and smaller gap have a better resolution.
However, more detailed investigation is necessary before drawing any firm conclusion.

For a numerical estimation of the energy resolution, we have followed the discussion in reference \cite{Energy1, Energy2}.
The intrinsic limit of the energy resolution in a gas detector is set by the statistical fluctuations in the primary number of electrons and the gain fluctuations.
The mean energy, measured after a repeated deposition of a fixed amount of energy in the detector, depends on the mean number of electron-ion pairs produced by the ionization $\bar{\mathrm{n}}$, the mean gas gain $\bar\mathrm{G}$ and the fraction of the electrons from the ionization reaching the amplification region i.e the electron transmission $\eta$. 

\begin{eqnarray}
\mathrm{E} = \bar{\mathrm{n}} \times \bar{\mathrm{G}} \times \eta \times \bar{\mathrm{f}}_{conv}
\end{eqnarray}

\noindent where $\bar{\mathrm{f}}_{conv}$ is the conversion factor, linking the number of charges to an energy value.
A contribution also comes from inhomogeneities of the parameters such as gas composition, pressure and temperature fluctuation, spatial inhomogeneities within the detector or the electronic noise.
But these contributions could of course be absorbed into other terms which are not considered in our calculation.

Applying error propagation analysis technique, the energy resolution R (the standard deviation of the distribution i.e $\sigma_\mathrm{E}/\mathrm{E}$) is thus simply the quadratic sum of the relative errors from these contributions.
 
\begin{eqnarray}
R = \frac{\sigma_\mathrm{E}}{\mathrm{E}} = \sqrt{\frac{\mathrm{F}}{\bar{\mathrm{n}}}~+~\frac{1}{\bar{\mathrm{n}}} \bigg(\frac{\sigma_{\mathrm{G}}}{\bar{\mathrm{G}}}\bigg)^2~+~\frac{(1-\eta)}{\eta\bar{\mathrm{n}}}}
\end{eqnarray}

\noindent where $\mathrm{F}$ is the Fano Factor and $\sigma_{\mathrm{G}}$ is the standard deviation of single avalanche.
In our simulation, we have considered the value of F to be 0.20 \cite{Chefdeville}.
The value of W (the mean energy per ion pair) has been estimated to be 26 eV, using the software Heed.   
As shown in Figure \ref{Energy-Drift-Figure}, the numerical results follow the experimental trend throughout the drift field range. 

\begin{figure}[hbt]
\centering
\subfigure[]
{\label{Energy-Pitch}\includegraphics[height=0.20\textheight]{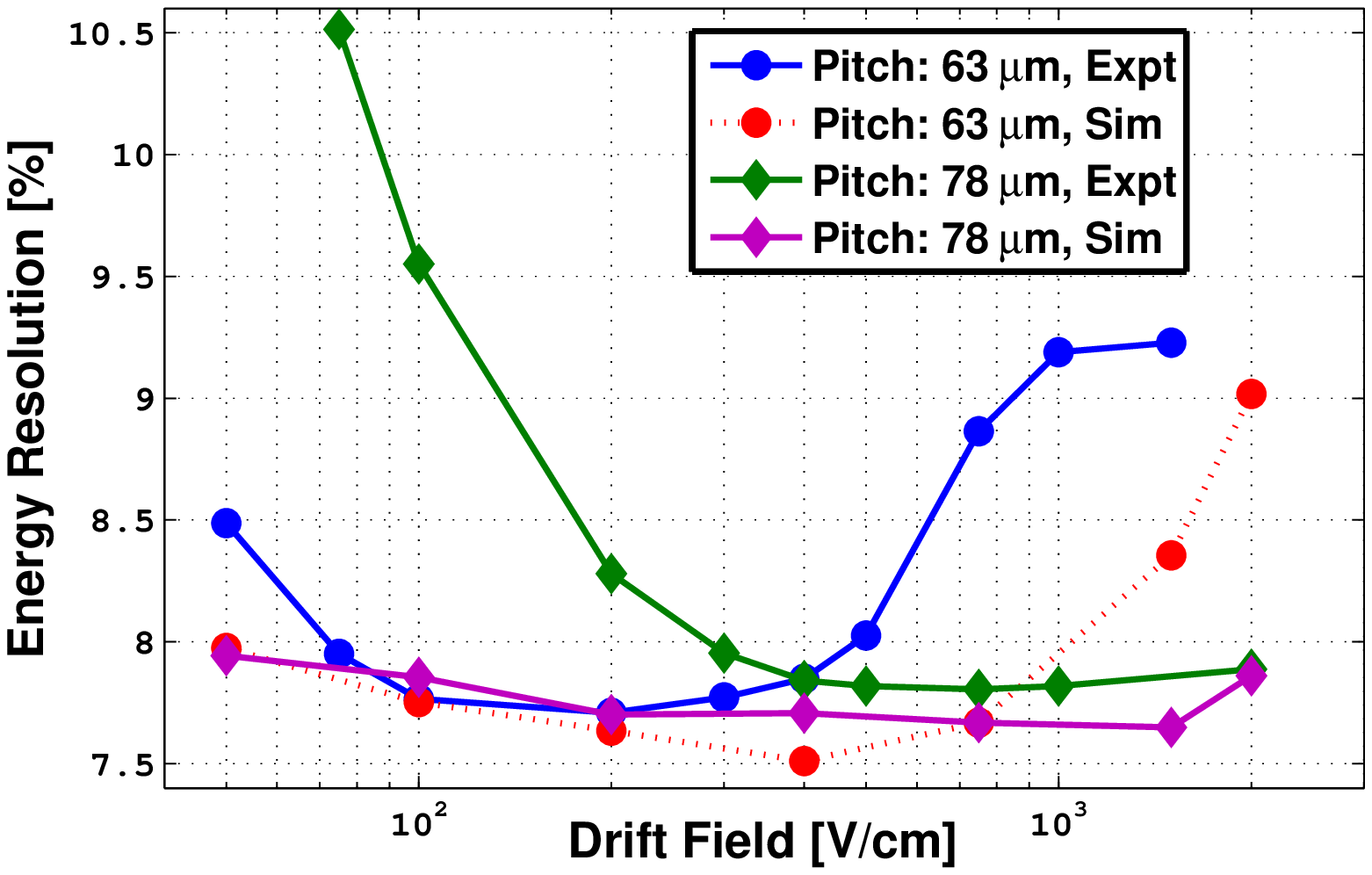}}
\subfigure[]
{\label{Energy-Gap}\includegraphics[height=0.20\textheight]{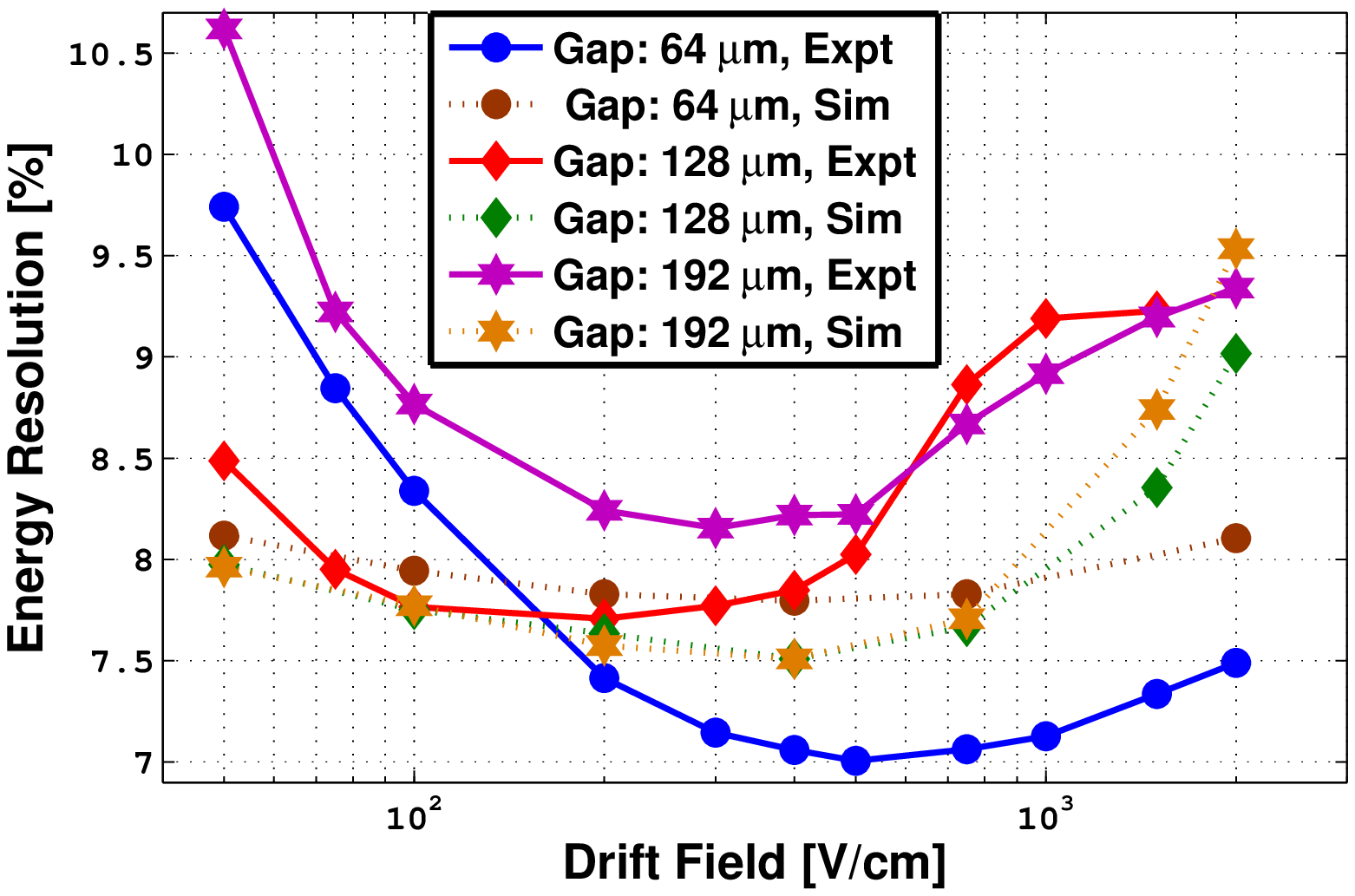}}
\caption{Variation of energy resolution with drift field in  Argon-Isobutane mixture (90:10) for (a) two different mesh hole pitches [pitch = $63~\mu\mathrm{m}$, mesh voltage = $-410~\mathrm{V}$, gain $\sim~3200$; pitch = $78~\mu\mathrm{m}$, mesh voltage = $-450~\mathrm{V}$, gain $\sim~4000$] having same amplification gap of $128~\mu\mathrm{m}$; (b) three different amplification gaps [gap = $64~\mu\mathrm{m}$, mesh voltage = $-330~\mathrm{V}$, gain $\sim~3000$; gap = $128~\mu\mathrm{m}$, mesh voltage = $-410~\mathrm{V}$, gain $\sim~3200$; gap = $192~\mu\mathrm{m}$, mesh voltage = $-540~\mathrm{V}$, gain $\sim~3600$] having same mesh hole pitch of $63~\mu\mathrm{m}$.} 
\label{Energy-Drift-Figure}
\end{figure}

\subsection{Ion Backflow}
\label{sec: ion back flow}

The operation of the gas avalanche detectors is often limited by the secondary effects, originating from the avalanche-induced photons and ions.
One of those secondary effects \cite{IBFPaul} is the ion backflow which is the drift of the positive ions produced in the avalanche, from the amplification region towards the drift mesh.
In TPCs operated at high particle rates, the space charge of the backflowing ions may modify the electron drift by locally disturbing the electric field.
The mesh of the Micromegas has the intrinsic property to naturally stop a large fraction of the secondary positive ions created in the avalanche.
Due to the field gradient between the drift and amplification regions and the periodic hole pattern, the field lines from the drift region are compressed in the vicinity of the micromesh holes and form a funnel having width of a few microns in the amplification region. 
As a result, an electron approaching the micromesh is focused towards the center of a hole and produces an avalanche inside the funnel. 
Due to the transverse diffusion, the avalanche also extends outside the funnel. 
Conversely, the ions, due to their larger mass, are not affected much by the diffusion and drift along the field lines. 
Assuming that they are emitted with the same distribution as the avalanche, most of these are collected by the micromesh and only a small fraction drifts back towards the drift mesh.

The ion backflow fraction can be defined as: 

\begin{eqnarray}
\mathrm{BF} = \frac {\mathrm{N}_\mathrm{b}} {\mathrm{N}_\mathrm{t}}
\end{eqnarray}

\noindent where $\mathrm{N}_\mathrm{t}$  is the average number of ions produced in an electron avalanche and $\mathrm{N}_\mathrm{b}$ the average number of the backflowing ions. 

Following the previous assumptions, the dependency of the ion backflow fraction on different detector parameters can be estimated through numerical simulation. In the two dimensional limit, it can be shown that,

\begin{eqnarray} \label{ibfeqn}
\mathrm{BF} \propto \frac {1} {\mathrm{FR}} \bigg(\frac {p} {\sigma_\mathrm{t}}\bigg)^2 
\end{eqnarray}

\noindent Here FR is the field ratio (amplification field/drift field) and p is the mesh pitch. 
$\sigma_\mathrm{t}$ is related to the transverse diffusion of the electron and given by $\mathrm{D}_{\mathrm{t}}\sqrt{\mathrm{z}}$ wherein $\mathrm{D}_{\mathrm{t}}$ is the transverse diffusion coefficient of electron and z is the path traversed \cite{Chefdeville}. 
  
The backflow fraction has been measured as:

\begin{eqnarray}
\mathrm{BF} =  \frac {(\mathrm{I}_\mathrm{C} - \mathrm{I}_\mathrm{P})} {(\mathrm{I}_\mathrm{M} + \mathrm{I}_\mathrm{C} - \mathrm{I}_\mathrm{P})}
\end{eqnarray}

\noindent where $\mathrm{I}_\mathrm{C}$ is the current measured on the first drift mesh (Figure \ref{SetupSaclay}) and is proportional to the number of ions collected on the drift mesh; 
$\mathrm{I}_\mathrm{P}$ is equal to the current measured on the first drift mesh without amplification and is referred to as the primary current; 
$\mathrm{I}_\mathrm{M}$ is the current measured in the micromesh and proportional to the number of ions collected on the mesh.
In our case, we ignore the primary ionization current, which is of the order of pico ampere.

Figure \ref{IBF-Expt-Sim} shows the variation of the measured ion backflow fraction with the field ratio for the $128~\mu\mathrm{m}$ bulk Micromegas detector (pitch $63~\mu\mathrm{m}$).
The use of the double drift mesh has clearly improved the results.
In this case, the experimental data points have been fitted using Eqn. \ref{ibfeqn} as shown in Figure \ref{IBF-Expt-Sim-Fit}.

For the numerical simulation, the electrons are injected in the drift region as described in the section \ref{sec: transparency}.
These electrons drift towards the amplification area where they get multiplied.
We have considered the primary ions in the drift region and the ions created in the avalanche for the simulation.
The backflow is calculated as 

\begin{eqnarray}
\mathrm{BF} = \frac {\mathrm{N}_\mathrm{id}} {({\mathrm{N}_\mathrm{id}}+{\mathrm{N}_\mathrm{im}})}
\end{eqnarray}

\noindent where ${\mathrm{N}_\mathrm{id}}$ is the number of ions collected at the drift plane and ${\mathrm{N}_\mathrm{im}}$ is the number of ions collected at the micromesh.
The simulation results agree quite well with the experimental data using double drift mesh and the fitted curve.

The dependence of the ion backflow fraction on the amplification gap and the mesh hole pitch has been investigated numerically and is shown in Figure \ref{IBF-Sim-DiffGapPitch}.
The bulk Micromegas detector with larger gap shows better performance while the backflow fraction is less for the smaller mesh hole pitch.
More work is already being carried out and will be reported soon.

\begin{figure}[hbt]
\centering
\subfigure[]
{\label{IBF-Expt-Sim}\includegraphics[height=0.205\textheight]{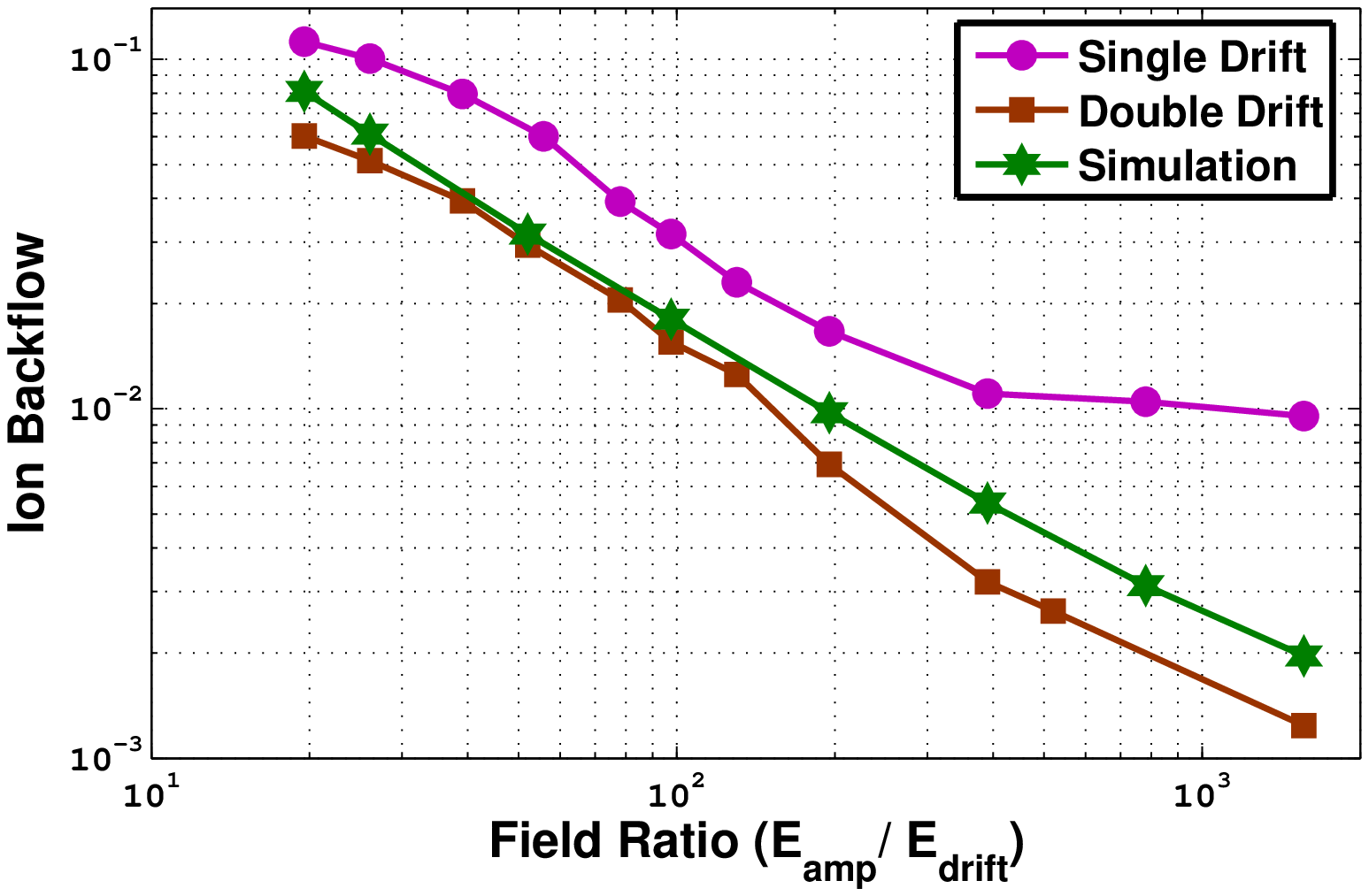}}
\subfigure[]
{\label{IBF-Expt-Sim-Fit}\includegraphics[height=0.205\textheight]{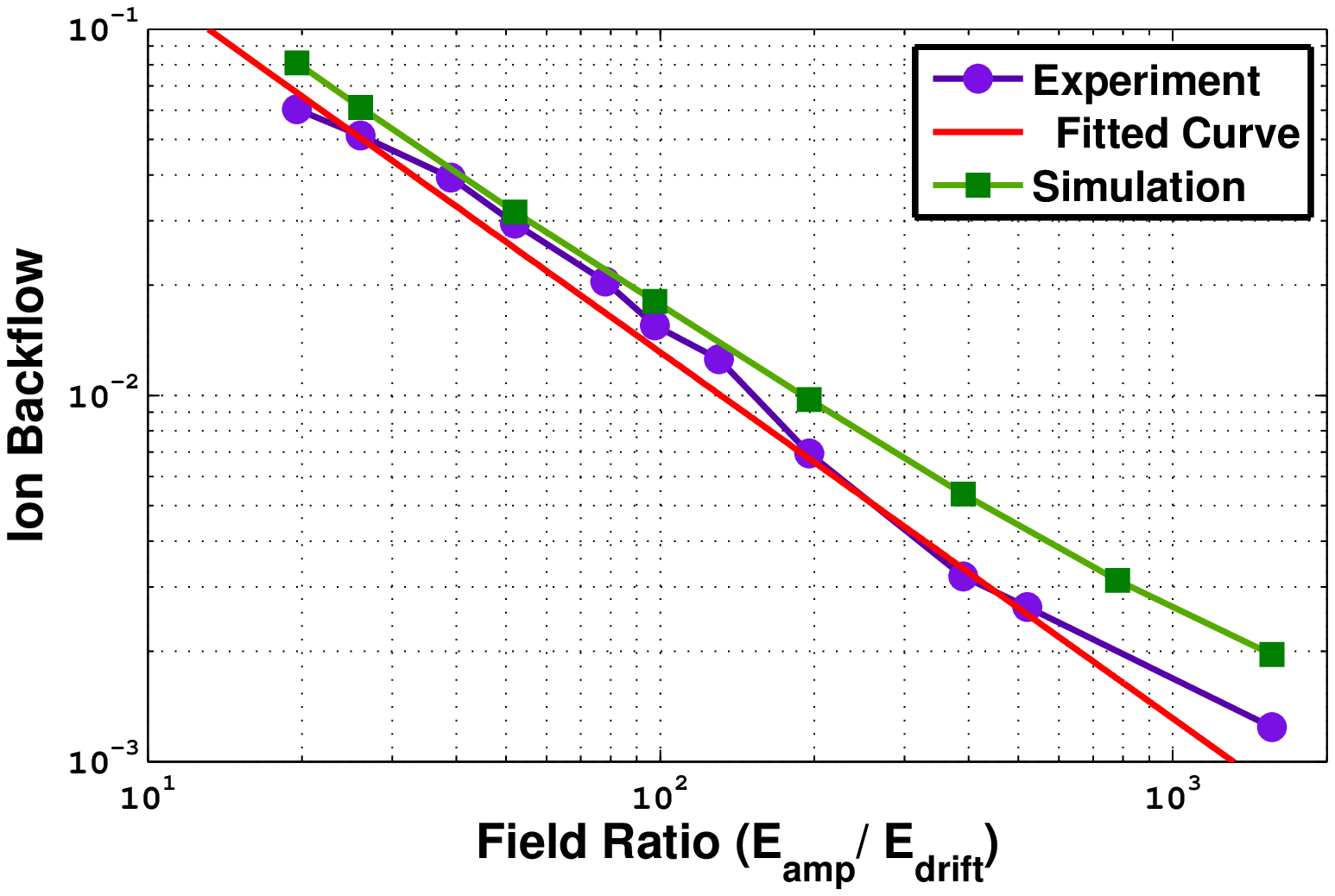}}
\subfigure[]
{\label{IBF-Sim-DiffGapPitch}\includegraphics[height=0.205\textheight]{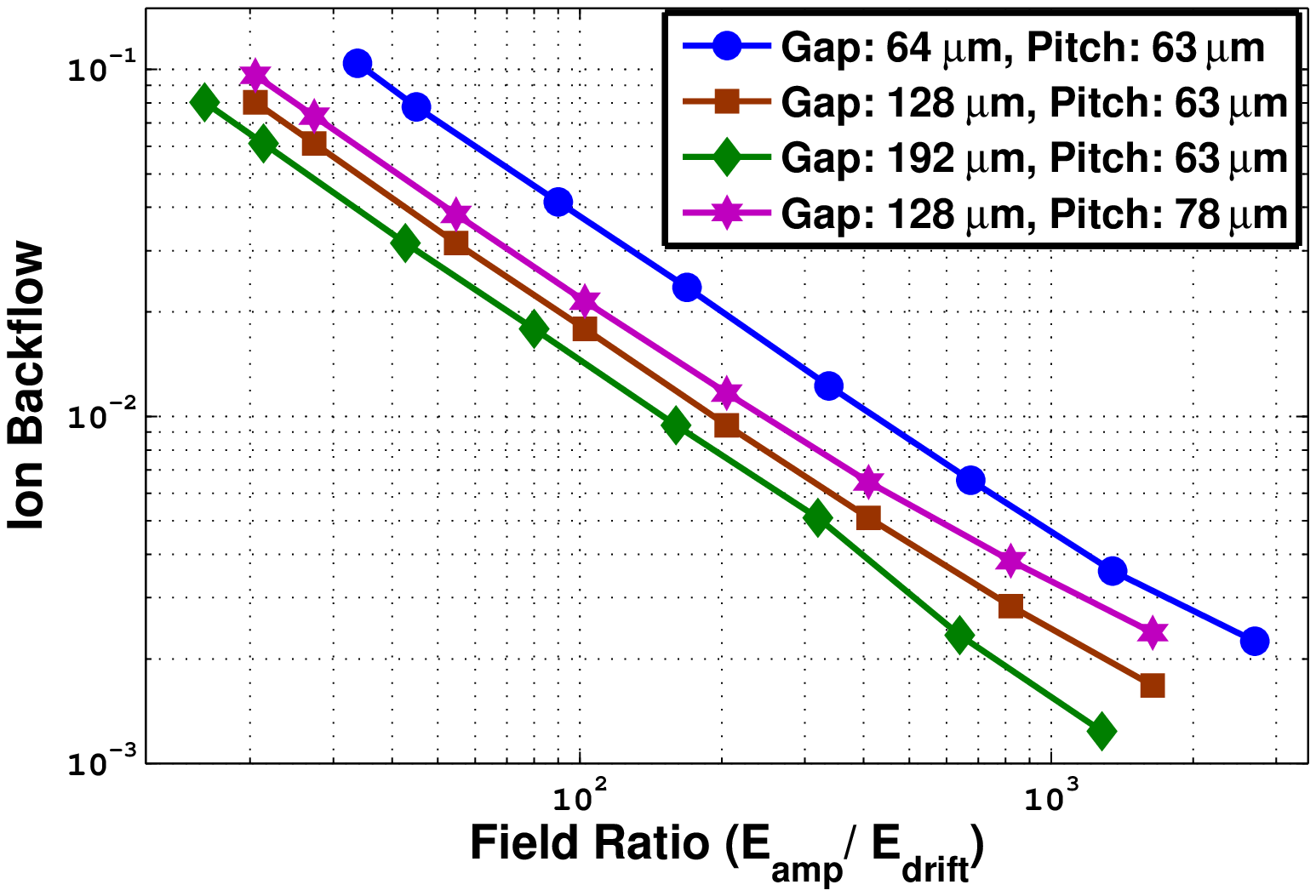}}
\caption{Variation of the ion backflow with the field ratio (amplification field/ drift field) in P10 gas; (a) comparison between experiment and simulation for $128~\mu\mathrm{m}$ amplification gap and $63~\mu\mathrm{m}$ pitch; (b) comparison with fitted curve using experimental data point and eqn 4.5; (c) numerical results for three different amplification gaps with same mesh hole pitch of $63~\mu\mathrm{m}$ and for two different pitch for a particular amplification gap of $128~\mu\mathrm{m}$. [gap = $64~\mu\mathrm{m}$, pitch = $63~\mu\mathrm{m}$, mesh voltage = $-410~\mathrm{V}$, gain = $\sim~2500$; gap = $128~\mu\mathrm{m}$, pitch = $63~\mu\mathrm{m}$, mesh voltage = $-500~\mathrm{V}$, gain = $\sim~2100$; gap = $192~\mu\mathrm{m}$, pitch = $63~\mu\mathrm{m}$, mesh voltage = $-620~\mathrm{V}$, gain = $\sim~2100$; gap = $128~\mu\mathrm{m}$, pitch = $78~\mu\mathrm{m}$, mesh voltage = $-550~\mathrm{V}$, gain = $\sim~2000$].}
\label{IBF-Results}
\end{figure}

\section{Conclusions}

In this paper, we present a comparative study between bulk Micromegas detectors having different amplification gaps and mesh hole pitches.
Various detector characteristics such as gain, electron transparency, energy resolution have been measured.
Experimentally, it has been observed that the bulk Micromegas with larger gap and smaller pitch shows better performance in terms of higher gain and lower backflow fraction at stable operating regime.
Comparisons of these measured data with the simulation results indicate that the physics of the amplifying micromesh-based structures (Micromegas) is quite well understood and suitably modeled mathematically.
The numerical estimates of the ion backflow fraction compare favorably with the measurements.

\section{Acknowledgment}

We thank our collaborators from the ILC-TPC collaboration for their help and suggestions.
We also thank Rui de Oliveira and the CERN MPGD workshop for technical support.
This work has partly been performed in the framework of the RD51 Collaboration. 
We happily acknowledge the help and suggestions of the members of the RD51 Collaboration.
We are thankful to Abhik Jash, Deb Sankar Bhattacharya, Wenxing Wang for their help in some measurements and Pradipta Kumar Das, Amal Ghoshal for their technical help.
We acknowledge CEFIPRA for partial financial support.
We thank the reviewer for providing useful suggestions.
Finally, we thank our respective Institutions for providing us with the necessary facilities.

\end{document}